\documentclass[aps,prl,preprint,groupedaddress]{revtex4-1}
\newcommand{\bea}{\begin{eqnarray}}
\newcommand{\eea}{\end{eqnarray}}
\newcommand{\be}{\begin{equation}}
\newcommand{\ee}{\end{equation}}

\newcommand{\dsl}{\pa \kern-0.5em /}

\newcommand{\pa}{\partial}

\begin{document}

\begin{flushright}

USTC-ICTS-18-14\\

\end{flushright}


\title{A distinctive signature of extra-dimensions: \\
The enhanced open string pair production}

\author{J. X. Lu}
\affiliation{Interdisciplinary Center for Theoretical Study\\
 University of Science and Technology of China, Hefei, Anhui
 230026, China}



\begin{abstract}
In analogy to the Schwinger pair production in QED,  there exists also the so-called open string pair production for a  system of two Dp branes, placed parallel at a separation, with at least one brane carrying a worldvolume electric flux, in Type II string theories. There is however no such pair production if  an isolated Dp brane carrying an electric flux is considered.  The produced open strings are directly related to the brane separation, therefore to the extra-dimensions if taken from the viewpoint of a brane observer. This pair production can be greatly enhanced if one Dp brane carries also a magnetic flux. The largest pair production rate occurs for $p = 3$, i.e., the D3 brane system, with the same applied fluxes.  A detection of this pair production by the brane observer as the charged particle/anti-charged particle pair one shall signal the existence of extra-dimensions and therefore provides a potential means to test the underlying string theories.\\
\end{abstract}

\maketitle

Whether there exist extra-dimensions remains an important yet unanswered question.  String theory, as a candidate of quantum gravity,  has built-in extra-dimensions and various non-perturbative solitonic extended objects such as Dp branes. One therefore expects string theory to provide means for exploring this question. In this work,  we discuss a special type of enhanced open string pair production, directly related to the extra-dimensions from the viewpoint of a brane observer, for a system of two D3 branes placed parallel at a separation and carrying certain worldvolume electric and magnetic fluxes. If we take one of D3 branes as our own 4-dimensional world, the brane observer, just like ourselves, knows about string theories but the observer can only, if all possible, detect the ends, not the whole, of the open strings so produced as charged particle/anti-charged particle pairs, in a fashion similar to the Schwinger pair production \cite{Schwinger:1951nm}, for example, by measuring the corresponding current in a laboratory setup within the brane. If this is indeed possible and the measurements against the applied electric and magnetic fluxes agree with the stringy prediction of the pair production rate, this then will have an implication on the existence of extra-dimensions, also a potential test of the underlying string theories.
   
 A static D3 brane in Type IIB superstring theory, being 1/2 Bogomol'ny-Prasad-Sommereld (BPS) vacuum-like object,  is stable. Its dynamics can also be described by a perturbative oriented open string with its two ends stick to the D3 brane along the transverse directions  \cite{Polchinski:1995mt} when the string coupling is small.  This open string is charge-neutral, having zero net-charge with its two ends carrying charge $+ 1$ and $- 1$, respectively. Just like the virtual electron/positron pair in quantum electrodynamics (QED) vacuum,  we have here the pair of virtual open string/anti open string, created from the present vacuum at some instant, existing for a short period of time, then annihilating to the vacuum. An observer on the brane can only sense the open string ends, not its whole, as virtual charged or anti-charged particles. So the pair of virtual open string/anti open string appears to the observer with one pair of their two nearby ends as the first pair of virtual charged particle/anti-charged particle and the other pair of their two other nearby ends as the second pair of virtual anti-charged particle/charged particle. So the quantum fluctuations from the perspective of brane observer are quite different from those of QED vacuum.

Just like the Schwinger pair production\cite{Schwinger:1951nm}, one would also expect to produce the charged particle/anti-charged particle or the open string pairs if a constant worldvolume electric field  is applied to an isolated D3 brane, depending on whether the observer is a brane one or a 10 dimensional bulk one.  However, in a sharp contrast, the stringy computations give a null result  due to the open strings being charge-neutral and their ends  experiencing the same electric field \cite{Bachas:1992bh, Lu:2017tnm}.  This is consistent with that a D3 carrying a constant electric field is a 1/2 BPS non-threshold bound state \cite{Lu:1999uca}, therefore being stable rather than unstable. The other way to understand this is as follows. To have the open string/anti open string pair detectable, they have to be separated infinitely away. However, this is impossible since either of them experiences a zero net-force under the action of the constant applied electric field less than its critical value.  In other words, there is no Schwinger-type pair production here.  
 
In order to have the pair production,  a possibility is to let the two ends of the charge-neutral open string experience different electric fluxes.  A simple setup for this is to consider two Dp branes placed parallel at a separation with each carrying a different electric flux  (we consider a general $p$ with $p = 3$ as a special case).  The open string pair production should then come from those virtual open strings with each connecting the two Dp branes along their transverse directions, therefore directly related to the extra-dimensions from the perspective of the brane observer.  Stringy computations do give a non-zero but usually vanishingly small rate for realistic electric fluxes applied \cite{Lu:2009yx}, due to the large string scale $M_{s} = 1/\sqrt{\alpha'}$ whose current constraint is from a few TeV upto the order of $10^{16} \sim 10^{17}$ GeV \cite{Berenstein:2014wva}.   This rate can however be greatly enhanced if at least one such Dp carries also a magnetic flux \cite{Lu:2009au, Lu:2017tnm}.  This enhancement makes it possible to detect the pair production and therefore 
to have the potential to address the question on the existence of extra-dimensions raised at the outset.

 We now compute this rate with the respective worldvolume dimensionless  flux $\hat F$ and $\hat F'$, both being antisymmetric $(p + 1)\times (p + 1)$ matrices with the same structure. For  the wanted enhancement, the non-vanishing components for $\hat F$ can be chosen, without loss of generality,  to be
 \be\label{flux}
 {\hat F}_{01} = - {\hat F}_{10} = - f, \qquad {\hat F}_{23} = - {\hat F}_{32} = -g,
 \ee
  with the electric flux $|f| < 1$  and the magnetic flux $ |g| < \infty$. We have the same for $\hat F'$ but denoting the corresponding fluxes each with a prime.  This choice of fluxes implies $p \ge 3$.  To have this rate, we first need to have the open string annulus interaction amplitude between the two Dp in its integral representation. This was given recently by the present author in \cite{Lu:2018suj} as,
 \be\label{annulusamplit}
\Gamma  = \frac{2^2 V_{p + 1} |f - f'||g - g'|}{ (8 \pi^2 \alpha')^{\frac{1 + p}{2}}} \int_0^\infty \frac{d t} {t^{\frac{p -1}{2}}} e^{- \frac{y^2 t}{2\pi\alpha' }} \frac{(\cosh \pi \nu'_0 t - \cos\pi \nu_0 t)^2}{\sin\pi\nu_0 t \sinh\pi\nu'_0 t} Z(t),
\ee
where
\be\label{zt}
Z(t) =  \prod_{n = 1}^\infty \frac{\prod^{2}_{i =1}\left[1 - 2 |z|^{2n} e^{ (- )^{i} \pi \nu'_0 t} \cos\pi\nu_0 t + |z|^{4n} e^{ (-)^{i} 2 \pi \nu'_0 t}\right]^{2}}{(1 - |z|^{2n})^4 \left(1 - 2 |z|^{2n} \cos2\pi \nu_0 t + |z|^{4n}\right) \left(1 - 2 |z|^{2n} \cosh2\pi \nu'_0 t + |z|^{4n}\right)}.
\ee
In the above, $|z| = e^{-\pi t} < 1$,  $y$ is the brane separation, $\alpha'$ the Regge slope parameter, and the electric parameter  $\nu_{0}\in [0, \infty)$ and the magnetic one $\nu'_{0}\in [0, 1]$ are determined by the electric fluxes and magnetic ones, respectively, as
\be\label{egparameter}
\tanh\pi\nu_{0} = \frac{|f - f'|}{1 - f f'}, \quad \tan \pi \nu'_{0} = \frac{|g - g'|}{1 + g g'}.
\ee  
The integrand in (\ref{annulusamplit}) has an infinite number of simple poles along the positive $t$-axis at $t_{k} = k/\nu_{0}$ with $k = 1, 2, \cdots$, for which $\sin \pi \nu_{0} t_{k} = 0$. These poles actually give rise to the decay of the underlying system via the so-called open string pair production.  The non-perturbative decay rate or usually also called pair production rate can be computed as the sum of the residues of the integrand at these poles times $\pi$ per unit worldvolume following  \cite{Bachas:1992bh} as 
  \be\label{eg3-pprate}
 {\cal W} = \frac{8\,   |f - f'||g - g'|}{(8\pi^2 \alpha')^{\frac{1 + p}{2}}} \sum_{k = 1}^\infty (-)^{k - 1} \left(\frac{\nu_0}{k}\right)^{\frac{p - 3}{2}} \frac{\left[\cosh\frac{\pi k \nu'_0}{\nu_0} - (-)^k\right]^2}{k \,\sinh \frac{\pi k \nu'_0}{\nu_0}} \, e^{- \frac{ k\, y^2}{2\pi \alpha' \nu_0}}\, Z (t_{k}),
\ee
where $ Z (t_{k})$, given by $Z (t)$ in (\ref{zt}) with $t = t_{k} = k/\nu_{0}$, takes its explicit expression as
\be\label{ztk}
Z (t_{k}) =   \prod_{n = 1}^\infty  \frac{\left[1 -  (-)^k \, e^{- \frac{2 n k \pi}{\nu_0} (1 - \frac{\nu'_0}{2 n})}\right]^4 \left[1 - (-)^k  \, e^{- \frac{2 n k \pi}{\nu_0}(1 + \frac{\nu'_0}{2 n})}\right]^4}{\left(1 - e^{- \frac{2 n k \pi}{\nu_0}}\right)^6 \left[1 -   \, e^{- \frac{2 n k \pi}{\nu_0} (1 - \nu'_0/ n)}\right] \left[1 -  \, e^{- \frac{2 n k \pi}{\nu_0}(1 + \nu'_0 / n)}\right]}.
\ee
Note that the odd and even $k$ in (\ref{eg3-pprate}) give their respective positive and negative contributions to the rate. For given electric and magnetic fluxes, this rate is highly suppressed by the brane separation $y$ and the integer $k$. We can qualitatively understand this by noting that the mass for each produced open string is  $ k\, T_{f} \, y $ with $T_{f} = 1/(2\pi \alpha')$ the fundamental string tension. So the larger  $k$ or $y$ or both are, the larger the mass is and therefore the more difficult the open string  can be produced.   For $f \neq f'$,  one can check that the larger $f$ or $f'$ is, the larger $\nu_{0}$ and $|f - f'|$ are and the larger the rate ${\cal W}$ is.  

In general, the presence of magnetic fluxes enhances this rate.  We here consider two special cases to show explicitly this enhancement. The first is the case of $g = g' \neq 0$ and we have  the enhancement from (\ref{eg3-pprate}) and (\ref{egparameter}) as 
\be
\frac{{\cal W}_{g = g' \neq 0}}{{\cal W}_0} = 1 + g^{2} > 1,
\ee
where  the zero-magnetic flux rate is \cite{Lu:2009yx} 
\be\label{zeromf}
{\cal W}_0 = \frac{32\, \nu_{0}\, |f - f'| }{(8 \pi^{2} \alpha')^{\frac{1 + p}{2}}} \sum_{l = 1}^{\infty} \frac{1}{(2 l - 1)^{2}} \left(\frac{\nu_{0}}{2 l - 1}\right)^{\frac{p - 3}{2}} e^{- \frac{ (2l - 1) \, y^2}{2\pi \alpha' \nu_0}}\prod^{\infty}_{n = 1} \left(\frac{1 + e^{- \frac{2n (2 l - 1) \pi}{\nu_{0}}}}{1 - e^{- \frac{2n (2 l - 1) \pi}{\nu_{0}}}}\right)^{8}.
\ee
A remark follows. From (\ref{zeromf}), one can check easily that ${\cal W}_{0} = 0$ if we set identical $f$ and $f'$ (now $\nu_{0} = 0$ from the first equation in (\ref{egparameter})).  This agrees with no Schwinger-type pair production of an isolated  D3 brane carrying a constant electric flux mentioned earlier. So to have the expected pair production, we need to have a nearby D3 brane in the transverse directions, which may be invisible (hidden or dark) to our own D3 brane.  

The second is the case of $\nu'_{0} /\nu_{0} \gg 1$.  This says  $\nu_{0} \ll 1$ since $\nu'_{0} \in (0, 1]$, implying $|f - f'| \ll 1$ from (\ref{egparameter}).  For a fixed $\nu'_{0} \in (0, 1]$ and a very small $\nu_{0}$,  the rate (\ref{eg3-pprate}) can be well approximated by its leading $k = 1$ term as 
\be\label{ega} 
{\cal W} \approx  \frac{4\,   |f - f'||g - g'|}{(8\pi^2 \alpha')^{\frac{1 + p}{2}}} \nu_0^{\frac{p - 3}{2}}  \, e^{- \frac{ y^2}{2\pi \alpha' \nu_0}} \, e^{\frac{\pi \nu'_{0}}{\nu_{0}}}.
\ee
The zero-magnetic flux rate (\ref{zeromf}) for the same small $\nu_{0}$ is now
\be\label{zmf}
{\cal W}_0\approx \frac{32\, \nu_{0}\, |f - f'| }{(8 \pi^{2} \alpha')^{\frac{1 + p}{2}}} \,  \nu_0^{\frac{p - 3}{2}}  \, e^{- \frac{  y^2}{2\pi \alpha' \nu_0}}.
\ee
The enhancement is then
\be\label{enhancement}
\frac{{\cal W}}{{\cal W}_0} = \frac{|g - g' |}{8\, \nu_{0}} \, e^{\frac{\pi \nu'_{0}}{\nu_{0}}},
\ee
which can be huge given that $\nu'_{0}/\nu_{0} \gg 1$ and $\nu_{0} \ll 1$. It has a value of $1.6 \times 10^{35}$, a very significant enhancement, for $\nu_0 = 0.02$ and $\nu'_0 = 0.5$ for a moderate choice of $g = - g' = 1$, noting $ g, g'  \in (- \infty, \infty) $ for $p = 3$. Note that the rate for $p > 3$ from (\ref{ega}) is  smaller than that for $p = 3$ by at least a factor of $(\nu_0/4 \pi)^{1/2} \approx 0.04$ for the above sample case. One may wonder if further enhancement can be achieved when we add an extra magnetic flux with similar structure. For example, for $p = 5$, we add a flux $\hat F_{45} = - \hat F_{54} = - \tilde g$ in addition to those given in (\ref{flux}). It turns out that this diminishes  rather than enhances the pair production rate.  The flux structure given in (\ref{flux}) actually gives the largest rate for each given $p \ge 3$  and moreover for the same applied fluxes the $p = 3$ rate is the largest  among these $p \ge 3$.   So this singles out the system of two D3 branes, therefore the 4-dimensional world. Curiously one of the D3 can be just our own world.  

Note that also for the $p = 3$ case, the string scale $\alpha'$ drops out, except for the exponential factor ${\rm exp} [- k y^{2}/(2\pi \alpha' \nu_{0})]$, for the rate  (\ref{eg3-pprate}) in practice for which the fluxes $f, f', g, g'$ are all very small (giving also very small $\nu_{0}$ and $\nu'_{0}$).  If we define a scale $m = T_{f} y = y/(2\pi \alpha')$, the aforementioned exponential factor depends only on  this scale and the $\alpha'$ also drops out.  One can check this easily if we set the dimensionless fluxes  $f = 2 \pi \alpha' \bar f, f' = 2\pi \alpha' \bar f', g = 2 \pi \alpha' \bar g, g' = 2 \pi \alpha' \bar g'$ with $\bar f, \bar f', \bar g, \bar g'$ the corresponding laboratory ones.  Now the rate  (\ref{eg3-pprate})  for $p = 3$ becomes 
\be\label{eg3pprate}
 {\cal W} = \frac{|\bar f - \bar f'||\bar g - \bar g'|}{2 \pi^2 } \sum_{k = 1}^\infty (-)^{k - 1} \frac{\left[\cosh\frac{\pi k \nu'_0}{\nu_0} - (-)^k\right]^2}{k \,\sinh \frac{\pi k \nu'_0}{\nu_0}} \, e^{- \frac{ k \pi m^{2}}{|\bar f - \bar f'| }},
\ee
where $\nu'_{0}/\nu_{0} = |\bar g - \bar g'|/|\bar f - \bar f'|$ and $Z (t_{k}) \approx 1$ for very small $\nu_{0}$ from (\ref{ztk})  has also been used. In practice, we can apply electric and magnetic fluxes 
only to our own D3 brane and have no control over the other D3. This amounts to setting, for example, $\bar f' = \bar g' = 0$, in (\ref{eg3pprate}). We have then, 
\be\label{eg3rate}
 {\cal W} = \frac{|\bar f||\bar g|}{2 \pi^2 } \sum_{k = 1}^\infty (-)^{k - 1} \frac{\left[\cosh\frac{\pi k \nu'_{0}}{\nu_0} - (-)^k\right]^2}{k \,\sinh \frac{\pi k \nu'_0}{\nu_0}} \, e^{- \frac{ k \pi m^{2}}{|\bar f| }},
\ee
with now $\nu'_{0}/\nu_{0} = |\bar g|/|\bar f|$. Given the alternative sign appearing in the sum, the present rate looks more like the scalar QED one \cite{Schwinger:1951nm, shewv, popov,iz, chop} than the spinor QED one \cite{shewv, bunkint, daughertyl}. According to \cite{nikishov}, the above rate should be more properly interpreted as the decay rate of the underlying system while the pair production rate is just the leading $k = 1$ term in (\ref{eg3rate}) since the higher $k$ correspond to more massive open strings, not the fundamental one.  With this, we now make a comparison of the present pair production rate with its correspondence in the spinor QED or scalar QED for the electric and magnetic fluxes specified. The present rate is
\be\label{eg3pprate-new}
{\cal W}^{(1)} = \frac{ 2 |\bar f||\bar g|}{(2 \pi)^2 } \frac{\left[\cosh\frac{\pi |\bar g|}{|\bar f|} +1\right]^2}{\sinh \frac{\pi |\bar g|}{|\bar f|}} \, e^{- \frac{  \pi m^{2}}{|\bar f| }},
\ee
the spinor QED rate \cite{nikishov} is
\be\label{spinorQED}
{\cal W}^{(1)}_{\rm spinor} = \frac{(q E) (q B)}{(2\pi)^{2}} \coth\left(\frac{\pi B}{E}\right) \, e^{- \frac{\pi m^{2}}{q E}},
\ee
and the scalar QED one \cite{nikishov}  is
\be\label{scalarQED}
{\cal W}^{(1)}_{\rm scalar} = \frac{(q E) (q B)}{2 (2\pi)^{2}} {\rm csch} \left(\frac{\pi B}{E}\right) \, e^{- \frac{\pi m^{2}}{q E}}.
\ee
In order to make comparisons, we need to identify $\bar f = q E$, $\bar g = q B$ and the present mass scale $m$ with the corresponding one in QED. The present rate (\ref{eg3pprate-new}) has similarities with but also important differences from the other two rates. Let us focus first on the pure electric case. We have now
\be\label{epprate}
{\cal W}^{(1)} =  8 \,{\cal W}^{(1)}_{\rm spinor} = 16\, {\cal W}^{(1)}_{\rm scalar} = \frac{8 (q E)^{2}}{4 \pi^{3}} \, e^{- \frac{\pi m^{2}}{q E}}.
\ee 
  The scalar QED rate is just half of the spinor QED one and this is due to the spinor factor $2 s + 1$. However, the present rate is 8 times of the spinor QED one and this can hardly be explained by the above spinor factor.  Recall that the present rate is a stringy one and the fluxes are taken care of non-linearly while the spinor or scalar rate is based on the corresponding linear field theory of QED.  This can be the source of the numerical difference. For example, the factor $[\cosh\pi |\tilde g|/|\tilde f| +1]^2$ in (\ref{eg3pprate-new}) contributes a factor of 4 for the pure electric case.  
 
 When the magnetic flux is turned on,   the present rate (\ref{eg3pprate-new}) is also always larger than the other two QED rates.  This is evident since 
 \be\label{rate-ratio}
 \frac{{\cal W}^{(1)}}{\,{\cal W}^{(1)}_{\rm spinor}} = \frac{2 \left[ 1 + \cosh \frac{\pi B}{E}\right]^{2}}{\cosh \frac{\pi B}{E}} > 1, \quad {\rm or } \quad  \frac{{\cal W}^{(1)}}{\,{\cal W}^{(1)}_{\rm scalar}} = 4 \left[1 +  \cosh\frac{ q B}{E}\right]^{2} > 1.
 \ee
 Especially when $B/E \gg 1$,   the rate  (\ref{eg3pprate-new}) is exponentially enhanced by the factor ${\rm exp} [\pi B/E]$ while the spinor rate (\ref{spinorQED}) has no such enhancement and the scalar rate (\ref{scalarQED}) on the contrast is exponentially suppressed by this factor.  Curiously all three rates have now the same numerical factor $1/(2 \pi)^{2}$. The above sharply different behavior between the rate  (\ref{eg3pprate-new}) and the spinor rate (\ref{spinorQED}) on magnetic flux lays a ground to distinguish the two when a detection of the underlying pair production becomes possible.   

If the two D3 brane separation is due to, for example, the standard model symmetry breaking, the mass scale $m$ should be naturally related to the symmetry breaking scale of a few hundred GeV.  This will then make its detection difficult and the only hope may be from LHC. On the other hand, if we interpret the other nearby D3 as invisible (hidden or dark) to our own D3, we usually don't have a priori knowledge of the mass scale $m$ for the rate (\ref{eg3pprate-new}).  If it happens to be on the order of electron mass, we then can test this rate with a tunable magnetic flux against the Schwinger pair production when the latter detection becomes feasible.  It is well-known that the lack of detection of Schwinger pair production up to now is due to the requirement of large constant electric field $E \sim 10^{18} \, {\rm V/m}$ which cannot be produced in laboratory.  The large enhancement with the presence of magnetic flux for our rate
 (\ref{eg3pprate-new}) can loosen this large field requirement to certain extent and may set such a detection sooner rather than later.  This can be even more true if the scale is smaller than that of electron mass. 
 
 Now there exist also various experiments involved large electromagnetic fields such as in relativistic heavy-ion collisions (RHIC), for example, $e E \sim e B \sim m^{2}_{\pi}$ at RHIC and $e E \sim e B \sim 10\, m^{2}_{\pi}$ at LHC.  However,  these large fields are the ones right after each collision but averaged to zero in a large event ensemble. As such, during the quark-gulon-plasma lifetime, it is so far still difficult to have a significant  detection of the underlying pair production. We hope that this situation can be improved soon and the aforementioned test can be carried out. 
 If the mass scale $m$ is on the order of $m_{\pi} = 140\, {\rm MeV}$, we may still use the RHIC to test the rate (\ref{eg3pprate-new}).   
 
 If a detection of the rate (\ref{eg3pprate-new}) from the perspective of a brane observer is indeed possible and the distinctive behavior of the rate against the applied tunable electric and magnetic fluxes is confirmed, their direct implication is the existence of extra dimensions and moreover this also gives a possible test  of the underlying string theories.


\begin{acknowledgments}
The author acknowledges support by grants from the NSF of China with Grant No: 11235010 and 11775212.
\end{acknowledgments}

\end{document}